\documentclass[a4paper,11pt]{article}

\usepackage{amsmath}
\usepackage{amssymb}
\usepackage{graphicx}
\usepackage{hyperref}
\usepackage[a4paper,left=0.99in, right=0.99in,top=1.1in, bottom=1.1in]{geometry}
\usepackage{color}
\usepackage[numbers]{natbib}

\usepackage{epsfig}
\usepackage{graphicx,color}

\newcommand{\be}{\begin{equation}}
\newcommand{\ee}{\end{equation}}

\usepackage{hyperref}

\usepackage{color}
\usepackage[normalem]{ulem}
\definecolor{pkcolor}{rgb}{0,0,1}
\definecolor{sscolor}{rgb}{1,0,1}
\definecolor{avhcolor}{rgb}{1,0,0}

\newcommand\pkout{\bgroup\markoverwith{\color{pkcolor}{\rule[0.4ex]{2pt}{0.8pt}}}\ULon}
\newcommand\avhout{\bgroup\markoverwith{\color{avhcolor}{\rule[0.4ex]{2pt}{0.8pt}}}\ULon}
\newcommand\ssout{\bgroup\markoverwith{\color{sscolor}{\rule[0.4ex]{2pt}{0.8pt}}}\ULon}

\title{ 
    \vskip 2cm
Small-$x$ dynamics in forward-central dijet decorrelations \\ at the LHC
}

\author{
A.~van Hameren,$^1$ P.~Kotko,$^1$ K.~Kutak,$^1$, S.~Sapeta$^2$\\\\
$^1$ {\small\it The H.\ Niewodnicza\'nski Institute of Nuclear Physics PAN}\\ {\small\it Radzikowskiego 152, 31-342 Krak\'ow, Poland}\\\\
$^2$ {\small\it CERN PH-TH, CH-1211, Geneva 23, Switzerland}\\
}

\date{}

\begin{document}
\maketitle

\vspace{-30em}
\begin{flushright}
  CERN-PH-TH-2014-070\\
  IFJPAN-IV-2014-6
\end{flushright}
\vspace{25em}

{\abstract
We provide a description, within the High Energy Factorization formalism, of
central-forward dijet decorrelation data measured by the CMS experiment and the predictions for nuclear modification ratio~$R_{pA}$ in p+Pb collisions. 
In our study, we use the unintegrated gluon density derived from the BFKL and BK
equations supplemented with subleading corrections and a hard
scale dependence.  The latter is introduced at the final step of the calculation
by reweighting the Monte Carlo generated events using suitable Sudakov form
factors, without changing the total cross section.
We achieve a good description of data in the whole region of the azimuthal
angle.
}

\section*{Introduction}

In Quantum Chromodynamics at high energies, there is a continuous search for
definite signatures of small-$x$ dynamics. Examples of expected signatures are:
high energy-enhanced rate of minijets between two hard jets that are far away in
rapidity~\cite{Mueller:1986ey}, suppression of rates of hadron production in
``dilute-dense'' scattering as compared to ``dilute-dilute''
scattering~\cite{Gribov:1984tu} or diffractive processes at high energies \cite{Flensburg:2012zy}.
Indeed, there have already been studies reporting evidence of small-$x$ effects for some observables 
in p+p and d+Au collisions \cite{Albacete:2010pg,Dusling:2012cg,Ducloue:2013bva}.

One of the best observables to study the dynamics at small-$x$ is the azimuthal
decorrelation, i.e.\ the differential distribution in the difference of the
azimuthal angle between two leading
jets~\cite{Vera:2007kn,Marquet:2003dm,Deak:2009xt,Deak:2010gk,Kutak:2012rf,vanHameren:2014lna,Nefedov:2013ywa}.
One of the reasons is that 
 small-$x$
 effects are inseparably related to the notion
of internal transverse momenta of (off-shell) gluons inside a hadron, which, according to
the 
Balitsky-Fadin-Kuraev-Lipatov (BFKL)
formalism 
\cite{Fadin:1975cb,Kuraev:1977fs,Balitsky:1978ic}, can be
large but cannot be arbitrarily 
small because of the importance of nonlinear effects
absent in the BFKL equation \cite{Kovchegov:1999yj,Kovchegov:1999ua,Balitsky:1995ub}. 
The internal
transverse momentum of a gluon can be viewed as a direct source of azimuthal
decorrelations, since it creates a jet momentum imbalance on the transverse plane.
On the other hand the decorrelations can be also made by parton showers by means of explicit additional emissions.
For the observables considered here, general purpose Monte Carlo
generators like $\mathtt{herwig}\textrm{++}$ \cite{Bahr:2008pv} and $\mathtt{pythia}$ \cite{Sjostrand:2007gs} were the only alternatives to describe
the data~\cite{CMS:2014oma}.

 With the present experimental program of CMS and ATLAS, one can test these
 effects 
 by 
investigating the pattern of radiation in a wide kinematical domain.
Of particular interest are jet observables, since in the factorization picture
of a collision,
the jets momenta are kinematically linked to momenta of initial state partons.
 Therefore,
scanning over a wide domain of jet $p_T$ allows one to test various physics
assumptions on the properties of gluon
distribution.

The effects that turn out to
play a crucial role are the kinematical effects enforcing momenta of the gluons
to be dominated by the transverse
components~\cite{Kwiecinski:1996td,Kwiecinski:1997ee} and the effects related to
soft gluon radiation of Sudakov type \cite{Sudakov:1954sw}. The need for such
effects in gluon cascades has been already recognized in
\cite{Catani:1989sg,Catani:1989yc,Kimber:1999xc} and has been formalized as the
Catani-Ciafaloni-Fiorani-Marchesini (CCFM) evolution equation or its nonlinear extension
\cite{Kutak:2011fu,Kutak:2012qk}. The latter takes into account saturation
effects. At present, the most commonly used framework providing a hard scale
dependence (i.e.\  Sudakov effects) in the parton density function is the 
Kimber-Martin-Ryskin (KMR)
evolution \cite{Kimber:1999xc,Kimber:2001sc}. In its final form, it is based on small-$x$ dynamics, angular
ordering, and it incorporates DGLAP effects.  Recently, it has been shown in
\cite{Mueller:2012uf}, for the case of color neutral particle production, that
the double logarithms of the type $\ln^2\left(\mu^2/k_T^2\right)$, where $\mu$
is the hard scale of the process and $k_T$ is the transverse momentum of the
initial state gluon, can be conveniently
resummed in the dipole approach, on top of small-$x$ resummation. 

\section*{Theoretical framework}

In the present work, we study central-forward dijet decorrelations using the High
Energy Factorization (HEF) approach
\cite{Gribov:1984tu, Catani:1990eg,Catani:1994sq,Collins:1991ty}.
In this framework, similarly to the standard collinear factorization, the cross
section is calculated as a convolution of a `hard
sub-process'~\cite{Lipatov:1995pn,Antonov:2004hh}~\footnote{For recent developments in hard matrix elements within HEF we refer the Reader to
\cite{vanHameren:2012if,vanHameren:2012uj,vanHameren:2013csa, Kotko:2014aba}.},
and nonperturbative parton densities, which take into account
longitudinal and transverse degrees of freedom. 
In practice, at low $x$, the gluons dominate over the quarks and therefore
one usually deals with the unintegrated gluon densities (UGDs) only.  We shall
describe different UGDs later in this section.

High Energy Factorization is obviously applicable only in a certain domain
and, in particular, one should be cautious about the following points: 
(i) although we use tree-level off-shell matrix elements and thus we
do not have explicit problems with factorization breaking by soft emissions, we
have to accept that the UGDs might not be universal,
(ii) below the  saturation scale, we should use more than just one
UGD,
but for the proton-proton collision studied here, the nonlinear effects
are rather weak. 
For more details concerning various factorization issues we refer to 
 \cite{Dominguez:2011wm,Xiao:2010sp,Dominguez:2011wm,Mulders:2011zt,Avsar:2011tz,Avsar:2012hj}.

In situations where the final state populates forward rapidity regions, one of
the longitudinal fractions of the hadron momenta is much smaller then the other,
$x_{A}\ll x_{B}$, and the following `hybrid' HEF formula is
used~\cite{Deak:2009xt}
\begin{multline}
d\sigma_{AB\rightarrow X}=\int \frac{d^{2}k_{TA}}{\pi}\int\frac{dx_{A}}{x_{A}}\,\int dx_{B}\, \\
\sum_{b}\mathcal{F}_{g^{*}/A}\left(x_{A},k_{TA},\mu \right)\, f_{b/B}\left(x_{B},\mu \right)\,
 d\hat{\sigma}_{g^{*}b\rightarrow X}\left(x_{A},x_{B},k_{TA},\mu \right),\label{eq:HENfact_2}
\end{multline}
where $\mathcal{F}_{g^{*}/A}$ is a UGD, $f_{b/B}$ are the collinear PDFs, and
$b$ runs over the gluon and all the quarks that can contribute to the production
of a multi-particle state $X$. 
Note that both $f_{b/B}$ and $\mathcal{F}_{g^{*}/A}$ depend on the hard
scale $\mu$.
 As we explain below, it is important to incorporate the hard scale dependence also in UGD.
The off-shell gauge-invariant matrix elements for multiple final states reside in $d\hat{\sigma}_{g^{*}b\rightarrow X}$. 
The condition $x_{B}\gg x_{A}$ is imposed by
proper cuts on the phase space of $X$.
It was shown in \cite{vanHameren:2013fla} that the phase space cuts for central-forward jets do
imply the aforementioned asymmetry condition.

In our computations, we used several different unintegrated gluon densities
$\mathcal{F}_{g^{*}/A} (x, k_{T},\mu)$:
\begin{itemize}
  \item
  The nonlinear KS (Kutak-Sapeta) unintegrated gluon
  density~\cite{Kutak:2012rf}, which comes from the extension of the BK
  (Balitsky-Kovchegov) equation \cite{Kutak:2003bd} following the prescription
  of KMS (Kwiecinski-Martin-Stasto) \cite{Kwiecinski:1997ee} to include
  kinematic constraint on the gluons in the chain, non-singular pieces of the
  splitting functions as well as contributions from sea quarks. The parameters
  of the gluon were set by the fit to $F_2$ data from HERA. This gluon can be
  determined for an arbitrary nucleus and, in the following, we shall use the KS
  densities for the proton and for lead.
  \item
  The linear KS gluon~\cite{Kutak:2012rf}, determined from the linearized
  version of the equation described above.
  \item
  The KMR hard scale dependent unintegrated gluon density~\cite{Kimber:1999xc,Kimber:2001sc}.
  It is obtained from the standard,
  collinear PDFs supplemented by the Sudakov form factor 
  and small-$x$ resummation of the BFKL type.
 The Sudakov form factor ensures no
  emissions between the scale of the gluon transverse momentum, $k_T$, and
  the scale of the hard process, $\mu$. The upper cutoff in the Sudakov form factor is
  chosen such that it imposes angular ordering in the last step of the
  evolution. The KMR gluon used in our study is based on MSTW 2008 LO~\cite{Martin:2009iq}.
  \item
  The standard collinear distribution $\mathcal{F}_{g^{*}} (x, k_{T}, \mu^2) =
  xg(x,\mu^2)\delta(k_T^2)$, which, when used in Eq.~(\ref{eq:HENfact_2}),
  reduces it to the collinear factorization formula. In this study we used the
  CTEQ10 NLO PDF set~\cite{Lai:2010vv}.

\end{itemize}

In addition, we supplement the KS linear and nonlinear UGDs with the Sudakov
resummation, which, as we shall see, turns out to be a necessary ingredient
needed to describe the data at 
moderate $\Delta\phi$.  The resummation is made on top of the Monte Carlo
generated events and it is motivated by the KMR prescription of the Sudakov form
factor. It effectively incorporates the dependence on a hard scale $\mu$ into the KS gluons, which by themselves do not exhibit such dependence.
A short description of the resummation model is
presented in the appendix.

\section*{Results}

\begin{figure}[t]
  \begin{center}
    \includegraphics[width=0.45\textwidth]{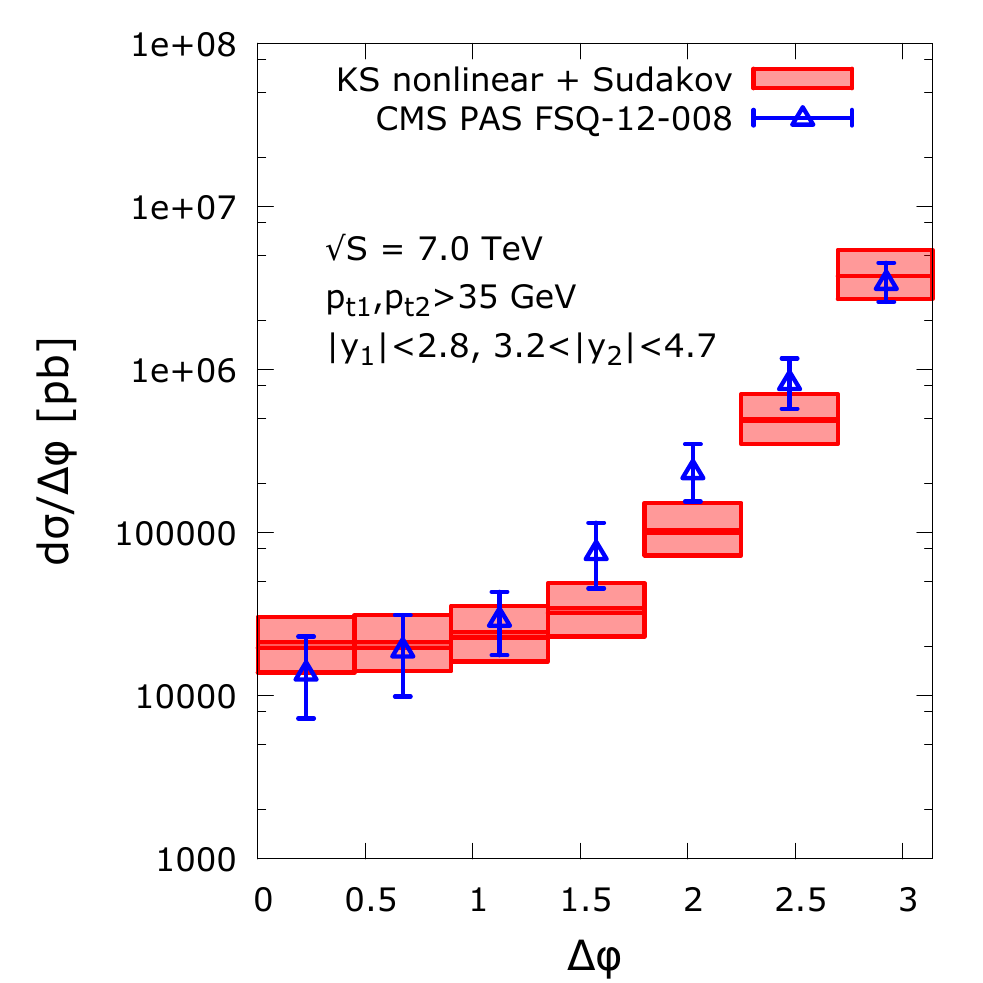}
    \includegraphics[width=0.45\textwidth]{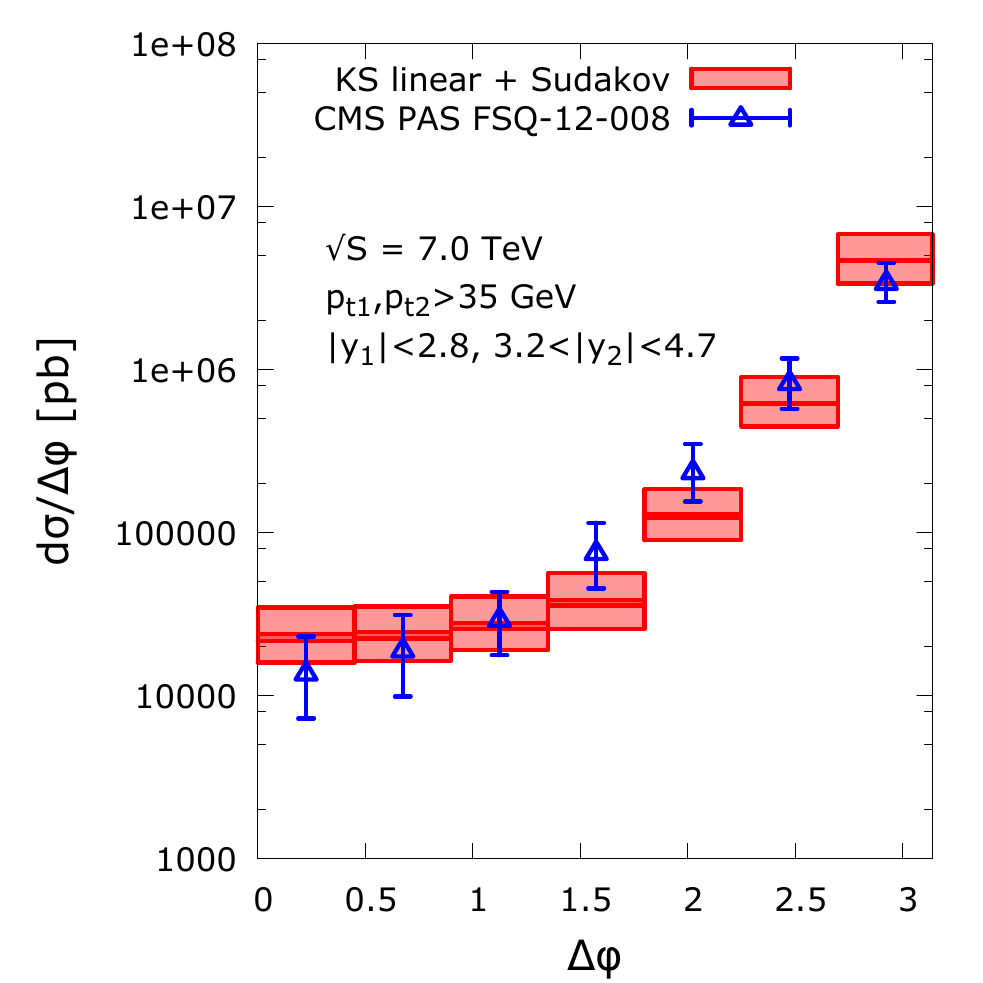} \\
    \includegraphics[width=0.45\textwidth]{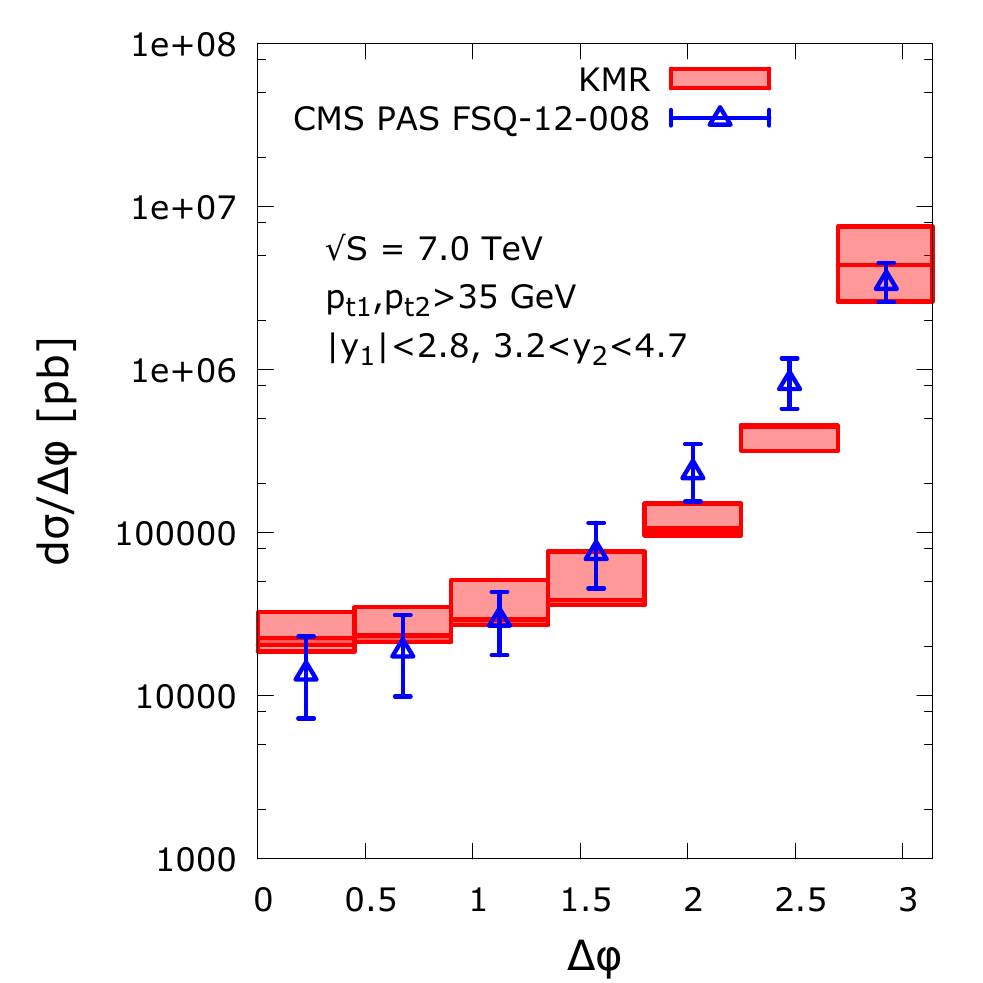} 
    \includegraphics[width=0.45\textwidth]{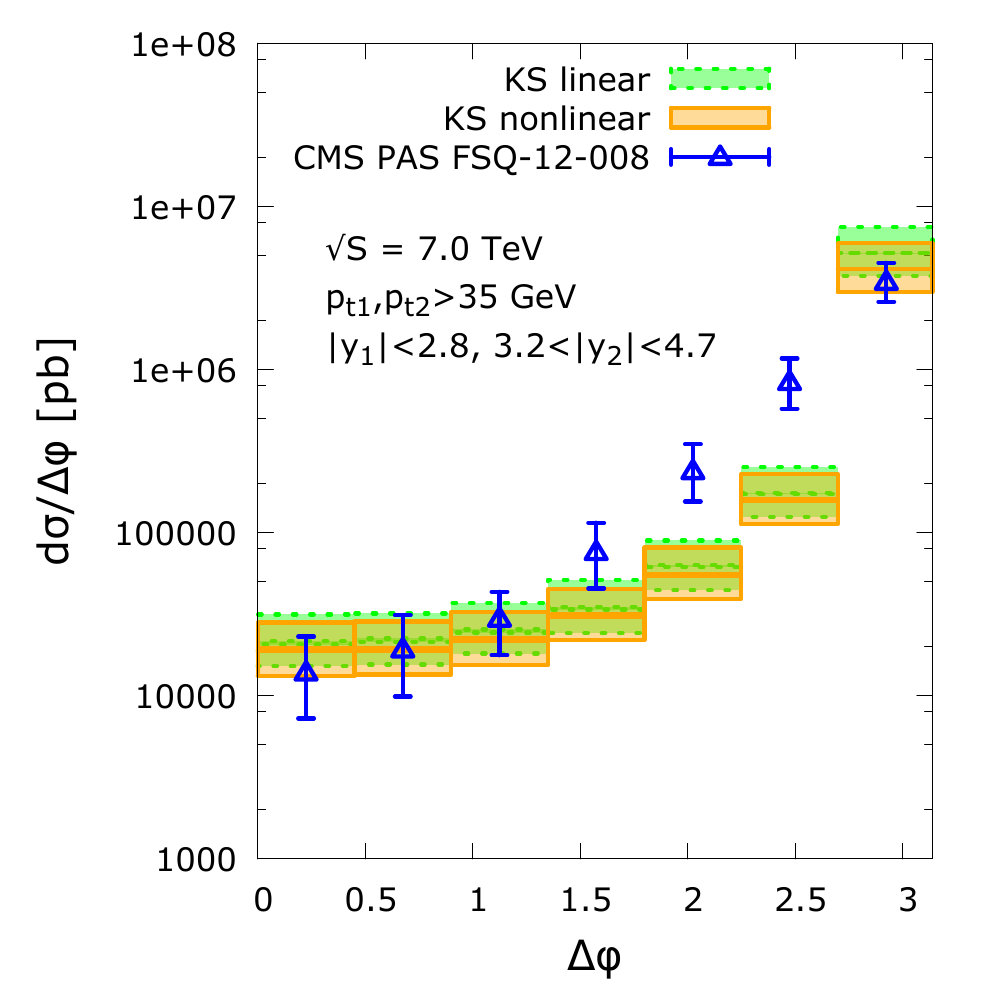} 
  \end{center}
  \caption{
  \small
  Comparison of CMS data for the inclusive dijet scenario with
  model predictions. 
  }
  \label{fig:veto_models_vs_CMS}
\end{figure}

\begin{figure}[t]
  \begin{center}
    \includegraphics[width=0.45\textwidth]{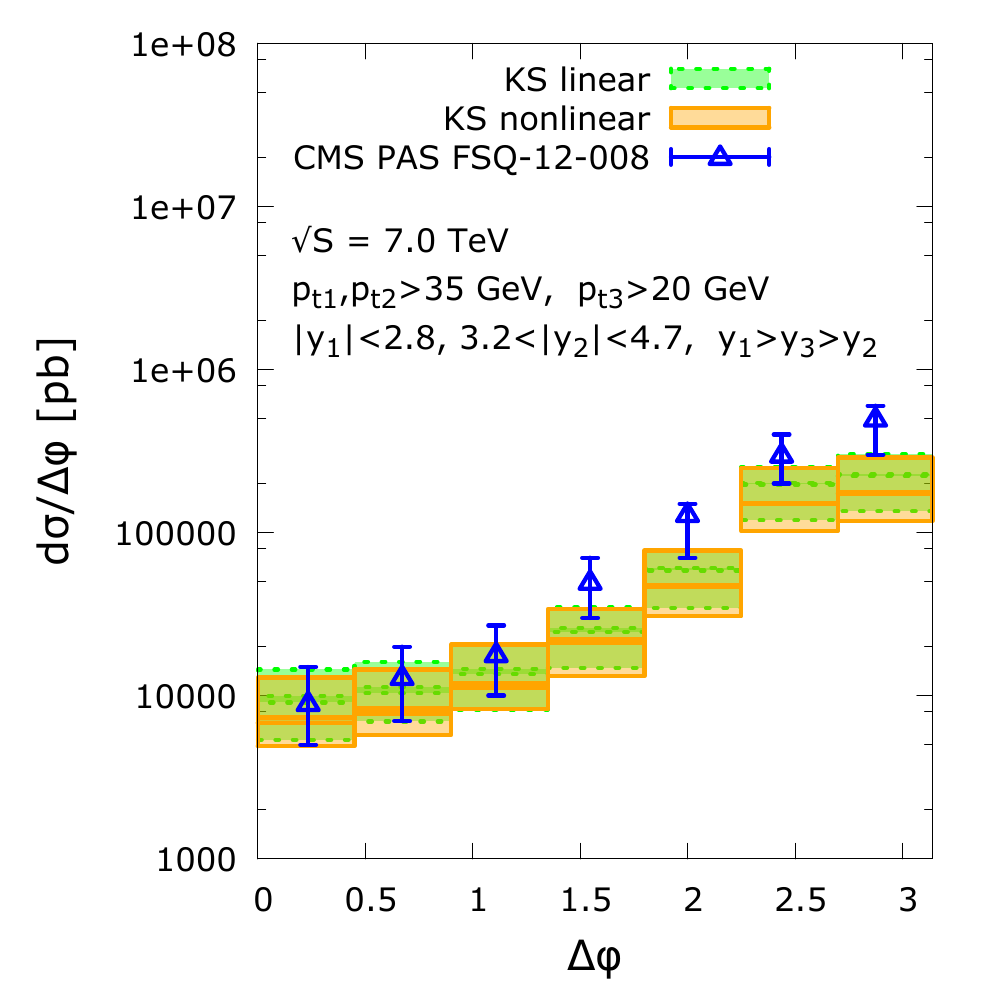}
    \includegraphics[width=0.45\textwidth]{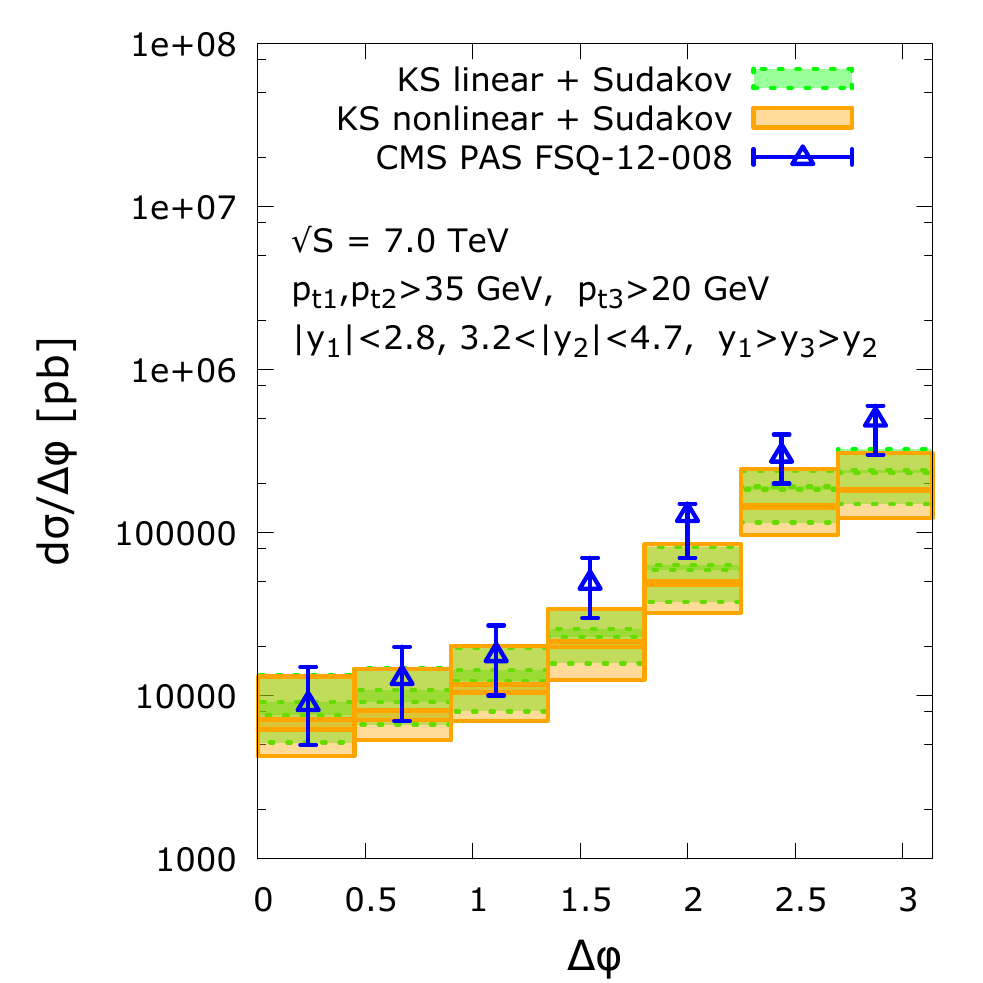} \\
    \includegraphics[width=0.45\textwidth]{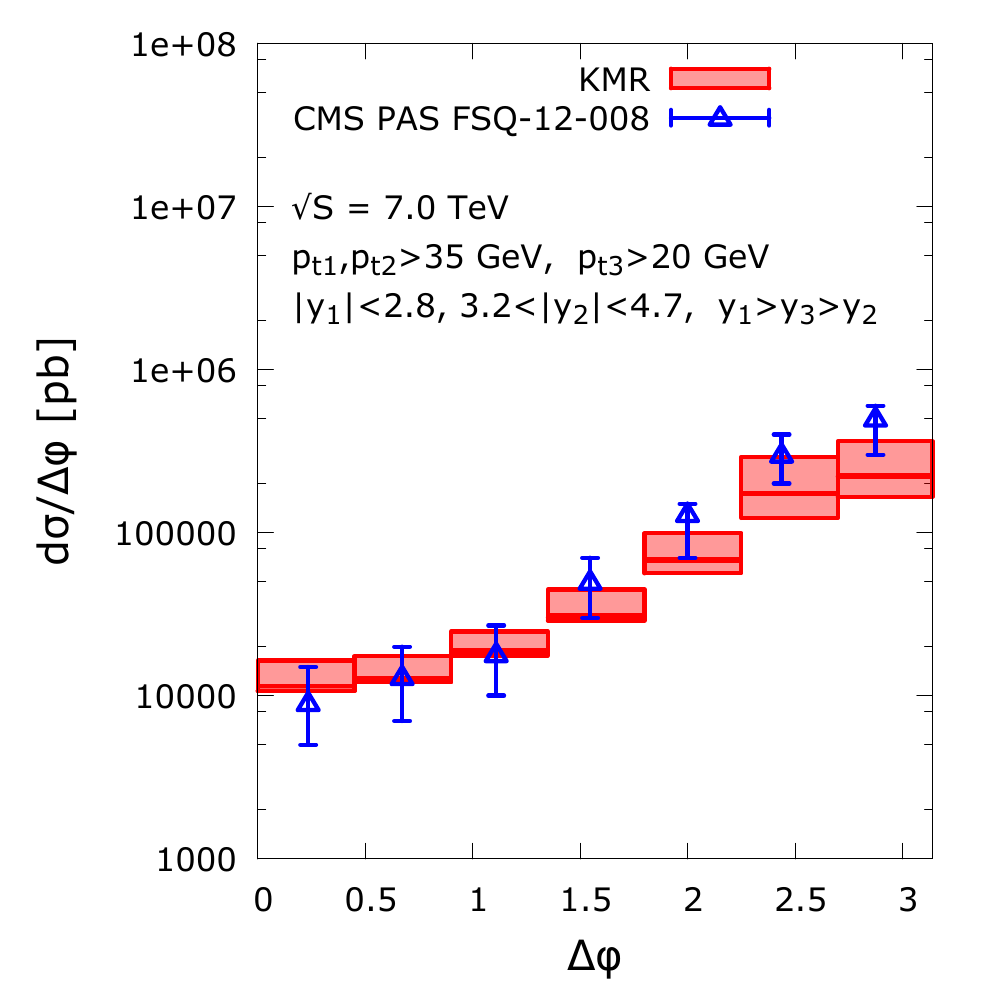} 
    \includegraphics[width=0.45\textwidth]{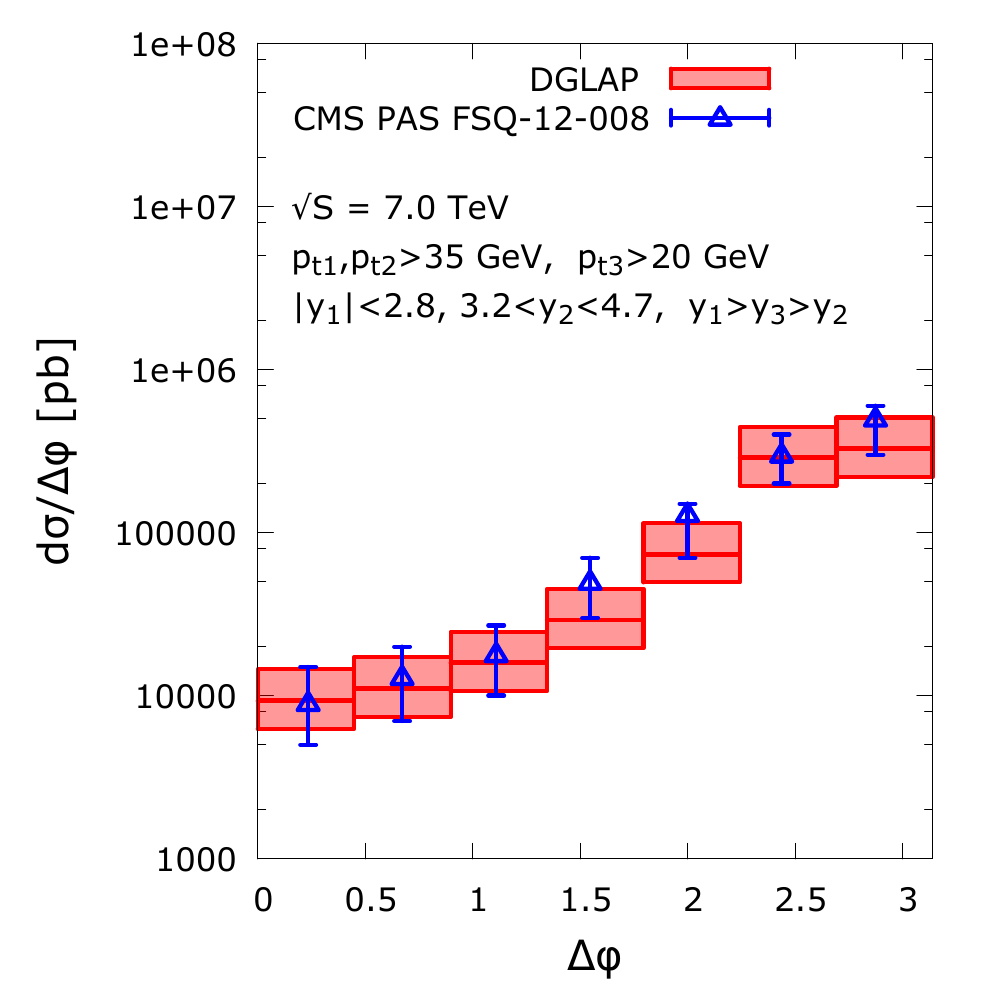} 
  \end{center}
  \caption{
  \small
  Comparison of CMS data for the dijet inside-jet-tag scenario with model 
  predictions. 
  }
  \label{fig:tag_models_vs_CMS}
\end{figure}

\begin{figure}
  \begin{center}
    \includegraphics[width=0.45\textwidth]{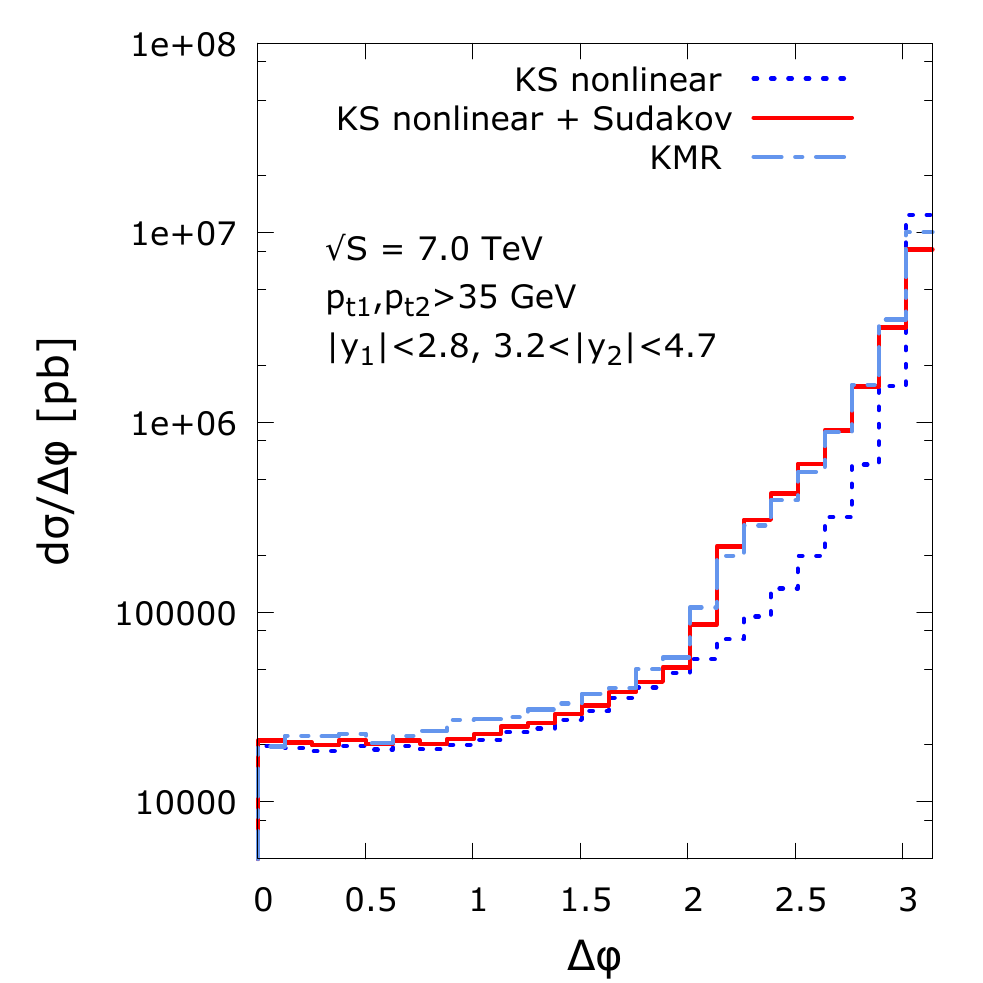} 
    \includegraphics[width=0.45\textwidth]{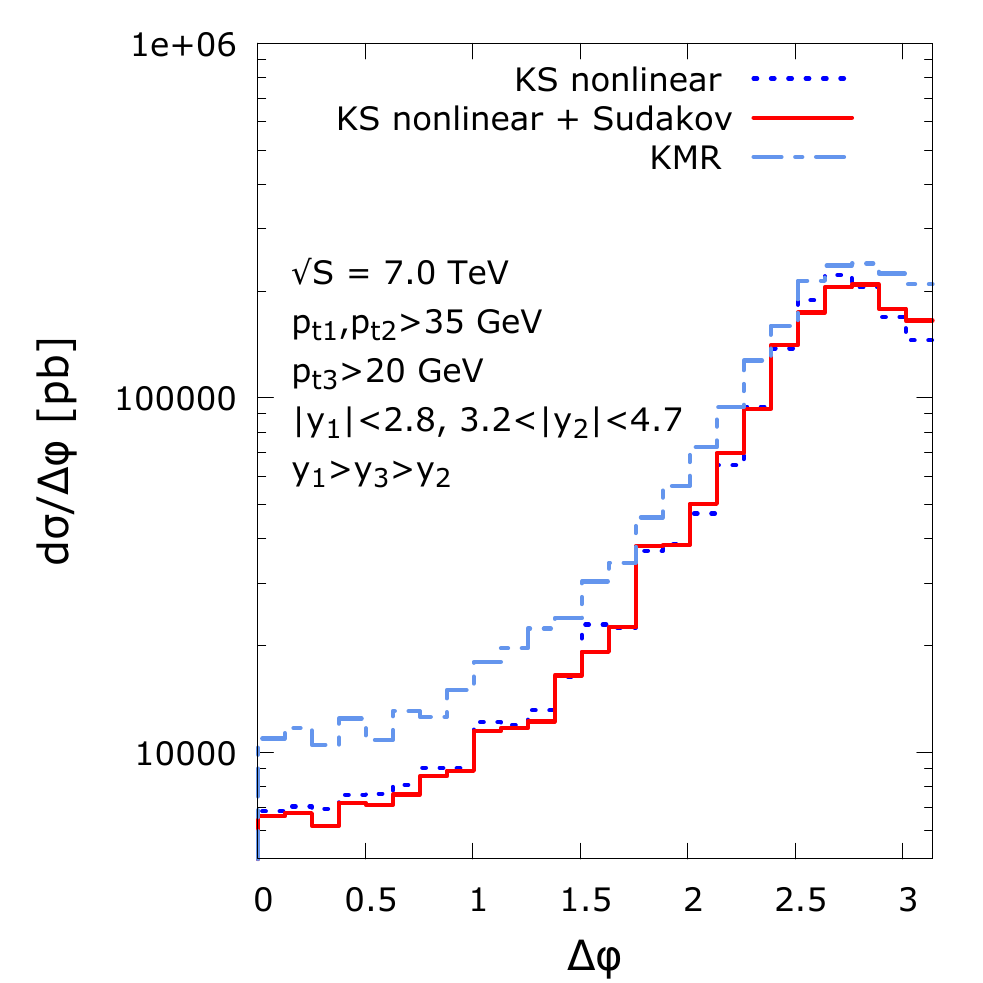} 
  \end{center}
  \caption{
  \small
   Illustration of the effect of the Sudakov resummation model on the KS
   nonlinear gluon density for inclusive dijet (left) and inside-jet-tag
   scenario (right). For comparison we plot also the KMR result.
  }
  \label{fig:models}
\end{figure}

\begin{figure}
  \begin{center}
    \includegraphics[width=0.45\textwidth]{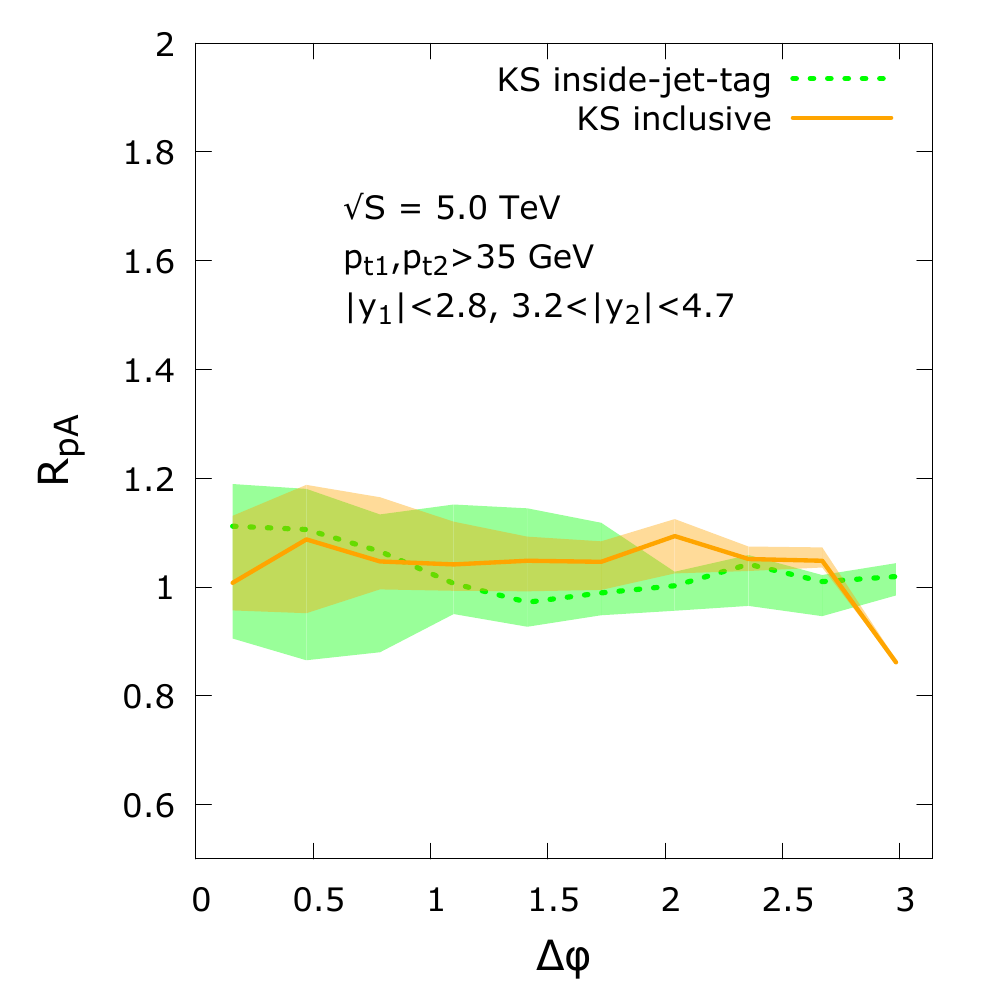}
    \includegraphics[width=0.45\textwidth]{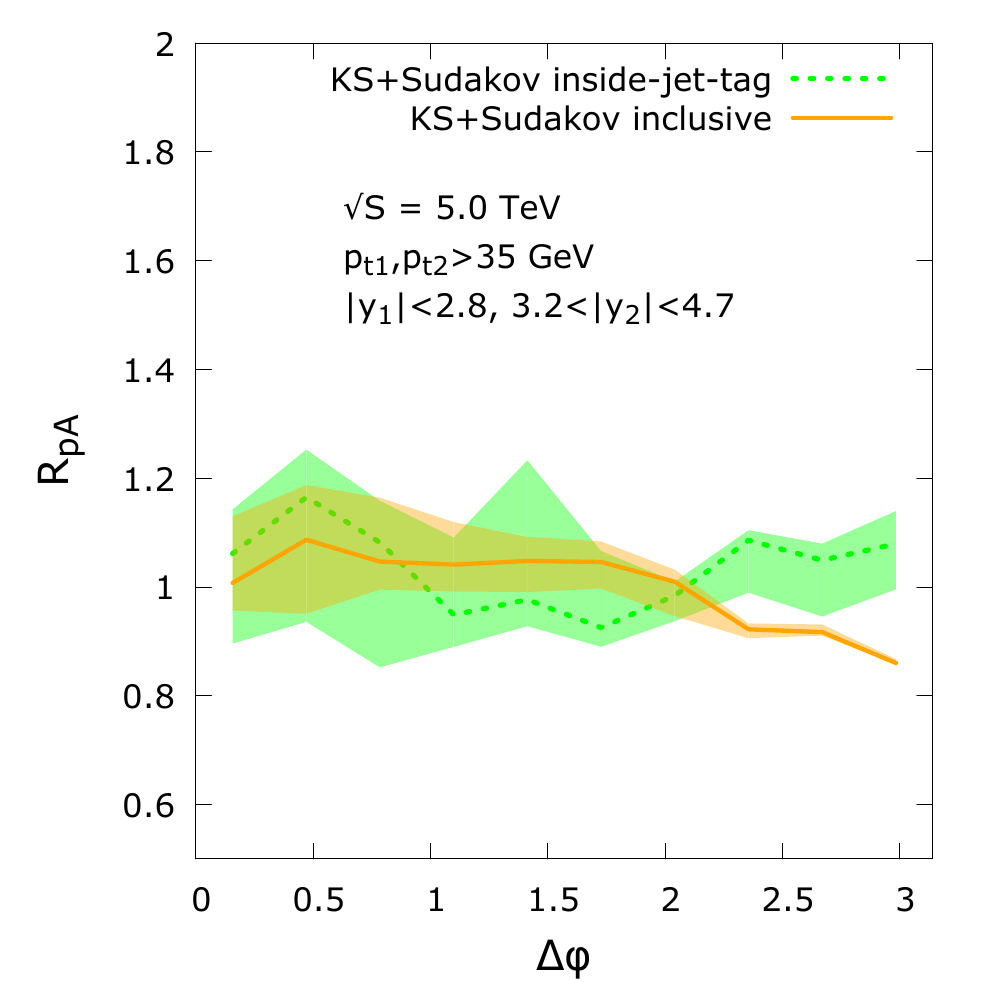}
  \end{center}
  \caption{
  \small
 Predictions for nuclear modification ratios for p+Pb collisions.
  }
  \label{fig:pPb}
\end{figure}

In this section, we present the results of our study of the azimuthal
decorrelations in the forward-central dijet production. Our framework enables
us to describe two scenarios considered in the  CMS forward-central dijet
measurement \cite{CMS:2014oma}: 
\begin{itemize}
\item 
\emph{Inclusive scenario}, which, in the experiment, corresponds to
selecting events with the two leading jets satisfying the cuts:  $p_{T,1,2} >
35\,$ GeV, $|y_1| < 2.8$, $3.2<|y_2|<4.9$ and with no extra requirement on 
further jets. These results are shown in Fig.~\ref{fig:veto_models_vs_CMS}.
\item
\emph{Inside-jet-tag scenario}, with the same selection on the two hardest jets
but, this time, a third jet with $p_T>20$ GeV is required between the forward
and the central region. The corresponding results are shown in
Fig.~\ref{fig:tag_models_vs_CMS}.
\end{itemize}
 
In Fig.~\ref{fig:veto_models_vs_CMS}, we present our 
results
 for the case of
the inclusive selection and compare them with the data from CMS. We show the
results obtained with the nonlinear and linear KS gluon, supplemented with the
Sudakov form factor (top left and top right, respectively), the KMR gluon
(bottom left) and an unmodified KS gluon (bottom right). We see that the
KS+Sudakov and KMR describe the data well. The error bands on the predictions
were obtained by varying the hard scale appearing explicitly in the running
coupling, UGDs, and the collinear PDFs by a factor $2^{\pm 1}$.
The calculations were performed independently by three programs:
$\mathtt{LxJet}$ \cite{Kotko_LxJet},  $\mathtt{forward}$~\cite{forward}, and a
program implementing the method of \cite{vanHameren:2012if}.

The results presented in Fig.~\ref{fig:veto_models_vs_CMS} provide evidence
in favour of the small-$x$ (or BFKL-like) dynamics. This dynamics produces gluon
emissions, unordered in $k_T$, which build up the non-vanishing $\Delta\phi$
distributions away from $\Delta\phi=\pi$. (A pure DGLAP based approach, without the
use of a parton shower, could of course only produce a delta function at
$\Delta\phi=\pi$.)
Furthermore, combining the above result with the recent analysis performed in
\cite{vanHameren:2014lna}, we conclude that the effects of higher
orders, like kinematical effects that allow for emissions at low $\Delta\phi$
are of crucial importance.
This alone is however not enough, since, as shown in
Fig.~\ref{fig:veto_models_vs_CMS}, one necessarily needs the Sudakov resummation
to improve the moderate $\Delta\phi$ (or equivalently moderate $k_T\sim 50\,
$ GeV) region.
These Sudakov effects are needed to resum virtual emissions between the hard
scale provided by the external probe and the scale of the emission from the gluonic
ladder. In other words, one has to assure that the external scale is the largest
scale in the scattering event.  Finally, Fig.~\ref{fig:veto_models_vs_CMS} indicates no need for saturation to describe
the azimuthal decorrelations in proton-proton collisions with these particular
experimental selection cuts,
although we see that it decreases the last bin at $\Delta\phi\sim\pi$
towards the data point.

Fig.~\ref{fig:tag_models_vs_CMS} shows the results obtained in the HEF approach
with the $2\rightarrow 3$ hard matrix elements. We also show the corresponding
result from pure DGLAP, i.e.\ with the HEF formula (\ref{eq:HENfact_2}) used in
the collinear limit.  We see that the linear and nonlinear KS results without
(top left) and with (top right) the Sudakov form factor nicely follow the experimental data
from the inside-jet-tag scenario. The description is also very good when the KMR
gluon is used (bottom left).  In the case of pure 
DGLAP calculation
(bottom right), the
3$^\text{rd}$ parton produced in the final state allows for generation of the
necessary transverse momentum imbalance between the two leading jets. This, in
turn, leads to a good description of the experimental data, even without the use
of a parton shower.

In Fig.~\ref{fig:models}, we compare the results obtained using
the KS gluon, with and without the Sudakov form factor, and the KMR gluon. We see that when
the Sudakov effects are included, the curves are almost on top of 
each other. This confirms the necessity of incorporating the hard scale dependence
to the unintegrated gluon densities. 
Further studies with in CCFM-based approaches~\cite{Jung:2010si,Deak:2010gk}
would be needed in order to get better understanding of these effects.

Finally, having described decorrelations data well, we are now ready for providing predictions for 
the nuclear modification ratios,
$R_{pA}$, in p+Pb collisions. 
They are defined as the ratios of p+Pb to p+p cross sections normalized to the number of nucleons.
In Fig.~\ref{fig:pPb}, we see that the suppression is more pronounced in the
inclusive scenario and further enhanced by the Sudakov effects. In particular we
see that inclusion of the Sudakov form factor changes the slope of the $R_{pA}$
ratio making the saturation effects visible in the wider range of the
$\Delta\phi$.

\section*{Acknowledgements}
We would like to thank P.~Cipriano, H.~Jung, G.~Gustafson, F.~Hautmann, A.~Knutsson, L.~Lonnblad, C.~Marquet, P.~van~Mechelen for many interesting and stimulating discussions.
We also would like to thank to R.~Maciula and A.~Szczurek for providing us the codes for the KMR gluon density. 
The work of A.~van~Hameren, K.~Kutak and P.~Kotko has been supported by Narodowe
Centrum Bada\'n i Rozwoju with grant LIDER/02/35/L-2/10/NCBiR/2011.

\appendix

\section*{Sudakov resummation model in Monte Carlo calculation}

The model is designed to supplement Monte Carlo generated events with Sudakov effects. 
It effectively incorporates an additional hard scale $\mu$ into the UGD and relies on two assumptions: 
(i) unitarity, i.e.\ the total cross section will not be affected, (ii) events with internal $k_T>\mu$
will be affected as little as possible.

Suppose we have a set of Monte Carlo generated events $\left(w_{i},X_{i}\right)$,
where $w_{i}$ is a weight and $X_{i}$ is a phase space point. An
observable is calculated according to
\begin{equation}
\mathcal{O}=\frac{\sigma}{W}\,\sum_{i}w_{i}F_{i}^{\mathcal{O}}\left(X_{i}\right),
\end{equation}
where $\sigma$ is the total cross section, $F_{i}^{\mathcal{O}}$
is a function defining the observable $\mathcal{O}$ (e.g.\ some combination
of step functions) and $W=\sum_{i}w_{i}$ is the total weight.
 Let us rewrite the observable $\mathcal{O}$ as follows
\begin{equation}
\mathcal{O}=\frac{\sigma}{W}\,\left[\sum_{i}w_{i}F_{i}^{\mathcal{O}}\left(X_{i}\right)\Theta\left(\mu_{i}>k_{Ti}\right)+\sum_{j}w_{j}F_{j}^{\mathcal{O}}\left(X_{j}\right)\Theta\left(k_{Tj}>\mu_{j}\right)\right].
\end{equation}
For each of the events from the first sum we incorporate the effect
of resummed unresolved emissions by including the Sudakov form factor $\Delta\left(\mu,k_T\right)$ which gives the probability that
a parton with transverse momentum $k_{T}$ will remain untouched while refining
the scale up to $\mu$. This is done be redefining the observable as follows
\begin{equation}
\widetilde{\mathcal{O}}=\frac{\sigma}{\widetilde{W}}\,\left[\sum_{i}w_{i}\Delta\left(\mu_{i},k_{Ti}\right)F_{i}^{\mathcal{O}}\left(X_{i}\right)\Theta\left(\mu_{i}>k_{Ti}\right)+\sum_{j}w_{j}F_{j}^{\mathcal{O}}\left(X_{j}\right)\Theta\left(k_{Tj}>\mu_{j}\right)\right]
\end{equation}
with
\begin{equation}
\widetilde{W}=\sum_{i}w_{i}\Delta\left(\mu_{i},k_{Ti}\right)\Theta\left(\mu_{i}>k_{Ti}\right)+\sum_{j}w_{j}\Theta\left(k_{Tj}>\mu_{j}\right).
\end{equation}
This changed the observable $\mathcal{O}\rightarrow\widetilde{\mathcal{O}}$.
Note, we do not change the total cross section, but we affect the events with $k_T>\mu$ contradicting our second assumption. 
This can be corrected by simply re-weighting them as follows
\begin{equation}
\overline{\mathcal{O}}=\frac{\sigma}{\overline{W}}\,\left[\sum_{i}w_{i}\Delta\left(\mu_{i},k_{Ti}\right)F_{i}^{\mathcal{O}}\left(X_{i}\right)\Theta\left(\mu_{i}>k_{Ti}\right)+\frac{\widetilde{W}}{W}\sum_{j}w_{j}F_{j}^{\mathcal{O}}\left(X_{j}\right)\Theta\left(k_{Tj}>\mu_{j}\right)\right]
\end{equation}
with $\overline{W}$ being the total weight as usual
\begin{equation}
\overline{W}=\sum_{i}w_{i}\Delta\left(\mu_{i},k_{Ti}\right)\Theta\left(\mu_{i}>k_{Ti}\right)+\frac{\widetilde{W}}{W}\sum_{j}w_{j}\Theta\left(k_{Tj}>\mu_{j}\right).
\end{equation}
As long as $\widetilde{W}/\overline{W}\sim 1$
our second assumption is satisfied. In case of the High Energy Factorization
and hard scale taken to be e.g.\ average $p_{T}$ of jets this is satisfied
easily. The events from the first sum will gain effectively the normalization
$W/\overline{W}>1$ relatively to the original observable $\mathcal{O}$ and account
for real emissions (a model of).

The Sudakov form factor appropriate for the above model can be constructed
along the lines given in \cite{Kimber:2001sc}
\begin{equation}
\Delta\left(\mu,k_{T}^{2}\right)=\exp\left(-\int_{k_{T}^{2}}^{\mu^{2}}\frac{dk_{T}^{'2}}{k{}_{T}^{'2}}\,\frac{\alpha_{s}\left(k{}_{T}^{'2}\right)}{2\pi}\,
\sum_{i} \int_{0}^{1-\epsilon\left(k_{T}^{'},\mu\right)}dz\,
P_{ig}\left(z\right)\right),\end{equation}
where $P_{ij}(z)$ are splitting functions, and 
$\epsilon\left(k_{T},\mu\right)=\frac{k_{T}}{\mu+k_{T}}$.

\bibliographystyle{apsrev}
\bibliography{small_x}

\begin{thebibliography}{53}
\expandafter\ifx\csname natexlab\endcsname\relax\def\natexlab#1{#1}\fi
\expandafter\ifx\csname bibnamefont\endcsname\relax
  \def\bibnamefont#1{#1}\fi
\expandafter\ifx\csname bibfnamefont\endcsname\relax
  \def\bibfnamefont#1{#1}\fi
\expandafter\ifx\csname citenamefont\endcsname\relax
  \def\citenamefont#1{#1}\fi
\expandafter\ifx\csname url\endcsname\relax
  \def\url#1{\texttt{#1}}\fi
\expandafter\ifx\csname urlprefix\endcsname\relax\def\urlprefix{URL }\fi
\providecommand{\bibinfo}[2]{#2}
\providecommand{\eprint}[2][]{\url{#2}}

\bibitem[{\citenamefont{Mueller and Navelet}(1987)}]{Mueller:1986ey}
\bibinfo{author}{\bibfnamefont{A.~H.} \bibnamefont{Mueller}} \bibnamefont{and}
  \bibinfo{author}{\bibfnamefont{H.}~\bibnamefont{Navelet}},
  \bibinfo{journal}{Nucl.Phys.} \textbf{\bibinfo{volume}{B282}},
  \bibinfo{pages}{727} (\bibinfo{year}{1987}).

\bibitem[{\citenamefont{Gribov et~al.}(1983)\citenamefont{Gribov, Levin, and
  Ryskin}}]{Gribov:1984tu}
\bibinfo{author}{\bibfnamefont{L.}~\bibnamefont{Gribov}},
  \bibinfo{author}{\bibfnamefont{E.}~\bibnamefont{Levin}}, \bibnamefont{and}
  \bibinfo{author}{\bibfnamefont{M.}~\bibnamefont{Ryskin}},
  \bibinfo{journal}{Phys.Rept.} \textbf{\bibinfo{volume}{100}},
  \bibinfo{pages}{1} (\bibinfo{year}{1983}).

\bibitem[{\citenamefont{Flensburg et~al.}(2012)\citenamefont{Flensburg,
  Gustafson, and Lönnblad}}]{Flensburg:2012zy}
\bibinfo{author}{\bibfnamefont{C.}~\bibnamefont{Flensburg}},
  \bibinfo{author}{\bibfnamefont{G.}~\bibnamefont{Gustafson}},
  \bibnamefont{and}
  \bibinfo{author}{\bibfnamefont{L.}~\bibnamefont{Lönnblad}},
  \bibinfo{journal}{JHEP} \textbf{\bibinfo{volume}{1212}}, \bibinfo{pages}{115}
  (\bibinfo{year}{2012}), \eprint{1210.2407}.

\bibitem[{\citenamefont{Albacete and Marquet}(2010)}]{Albacete:2010pg}
\bibinfo{author}{\bibfnamefont{J.~L.} \bibnamefont{Albacete}} \bibnamefont{and}
  \bibinfo{author}{\bibfnamefont{C.}~\bibnamefont{Marquet}},
  \bibinfo{journal}{Phys.Rev.Lett.} \textbf{\bibinfo{volume}{105}},
  \bibinfo{pages}{162301} (\bibinfo{year}{2010}), \eprint{1005.4065}.

\bibitem[{\citenamefont{Dusling and Venugopalan}(2013)}]{Dusling:2012cg}
\bibinfo{author}{\bibfnamefont{K.}~\bibnamefont{Dusling}} \bibnamefont{and}
  \bibinfo{author}{\bibfnamefont{R.}~\bibnamefont{Venugopalan}},
  \bibinfo{journal}{Phys.Rev.} \textbf{\bibinfo{volume}{D87}},
  \bibinfo{pages}{051502} (\bibinfo{year}{2013}), \eprint{1210.3890}.

\bibitem[{\citenamefont{Ducloué et~al.}(2014)\citenamefont{Ducloué,
  Szymanowski, and Wallon}}]{Ducloue:2013bva}
\bibinfo{author}{\bibfnamefont{B.}~\bibnamefont{Ducloué}},
  \bibinfo{author}{\bibfnamefont{L.}~\bibnamefont{Szymanowski}},
  \bibnamefont{and} \bibinfo{author}{\bibfnamefont{S.}~\bibnamefont{Wallon}},
  \bibinfo{journal}{Phys.Rev.Lett.} \textbf{\bibinfo{volume}{112}},
  \bibinfo{pages}{082003} (\bibinfo{year}{2014}), \eprint{1309.3229}.

\bibitem[{\citenamefont{Sabio~Vera and Schwennsen}(2007)}]{Vera:2007kn}
\bibinfo{author}{\bibfnamefont{A.}~\bibnamefont{Sabio~Vera}} \bibnamefont{and}
  \bibinfo{author}{\bibfnamefont{F.}~\bibnamefont{Schwennsen}},
  \bibinfo{journal}{Nucl.Phys.} \textbf{\bibinfo{volume}{B776}},
  \bibinfo{pages}{170} (\bibinfo{year}{2007}), \eprint{hep-ph/0702158}.

\bibitem[{\citenamefont{Marquet and Peschanski}(2004)}]{Marquet:2003dm}
\bibinfo{author}{\bibfnamefont{C.}~\bibnamefont{Marquet}} \bibnamefont{and}
  \bibinfo{author}{\bibfnamefont{R.~B.} \bibnamefont{Peschanski}},
  \bibinfo{journal}{Phys.Lett.} \textbf{\bibinfo{volume}{B587}},
  \bibinfo{pages}{201} (\bibinfo{year}{2004}), \eprint{hep-ph/0312261}.

\bibitem[{\citenamefont{Deak et~al.}(2009)\citenamefont{Deak, Hautmann, Jung,
  and Kutak}}]{Deak:2009xt}
\bibinfo{author}{\bibfnamefont{M.}~\bibnamefont{Deak}},
  \bibinfo{author}{\bibfnamefont{F.}~\bibnamefont{Hautmann}},
  \bibinfo{author}{\bibfnamefont{H.}~\bibnamefont{Jung}}, \bibnamefont{and}
  \bibinfo{author}{\bibfnamefont{K.}~\bibnamefont{Kutak}},
  \bibinfo{journal}{JHEP} \textbf{\bibinfo{volume}{0909}}, \bibinfo{pages}{121}
  (\bibinfo{year}{2009}), \eprint{0908.0538}.

\bibitem[{\citenamefont{Deak et~al.}(2010)\citenamefont{Deak, Hautmann, Jung,
  and Kutak}}]{Deak:2010gk}
\bibinfo{author}{\bibfnamefont{M.}~\bibnamefont{Deak}},
  \bibinfo{author}{\bibfnamefont{F.}~\bibnamefont{Hautmann}},
  \bibinfo{author}{\bibfnamefont{H.}~\bibnamefont{Jung}}, \bibnamefont{and}
  \bibinfo{author}{\bibfnamefont{K.}~\bibnamefont{Kutak}}
  (\bibinfo{year}{2010}), \eprint{1012.6037}.

\bibitem[{\citenamefont{Kutak and Sapeta}(2012)}]{Kutak:2012rf}
\bibinfo{author}{\bibfnamefont{K.}~\bibnamefont{Kutak}} \bibnamefont{and}
  \bibinfo{author}{\bibfnamefont{S.}~\bibnamefont{Sapeta}},
  \bibinfo{journal}{Phys.Rev.} \textbf{\bibinfo{volume}{D86}},
  \bibinfo{pages}{094043} (\bibinfo{year}{2012}), \eprint{1205.5035}.

\bibitem[{\citenamefont{van Hameren et~al.}(2014)\citenamefont{van Hameren,
  Kotko, Kutak, Marquet, and Sapeta}}]{vanHameren:2014lna}
\bibinfo{author}{\bibfnamefont{A.}~\bibnamefont{van Hameren}},
  \bibinfo{author}{\bibfnamefont{P.}~\bibnamefont{Kotko}},
  \bibinfo{author}{\bibfnamefont{K.}~\bibnamefont{Kutak}},
  \bibinfo{author}{\bibfnamefont{C.}~\bibnamefont{Marquet}}, \bibnamefont{and}
  \bibinfo{author}{\bibfnamefont{S.}~\bibnamefont{Sapeta}}
  (\bibinfo{year}{2014}), \eprint{1402.5065}.

\bibitem[{\citenamefont{Nefedov et~al.}(2013)\citenamefont{Nefedov, Saleev, and
  Shipilova}}]{Nefedov:2013ywa}
\bibinfo{author}{\bibfnamefont{M.}~\bibnamefont{Nefedov}},
  \bibinfo{author}{\bibfnamefont{V.}~\bibnamefont{Saleev}}, \bibnamefont{and}
  \bibinfo{author}{\bibfnamefont{A.~V.} \bibnamefont{Shipilova}},
  \bibinfo{journal}{Phys.Rev.} \textbf{\bibinfo{volume}{D87}},
  \bibinfo{pages}{094030} (\bibinfo{year}{2013}), \eprint{1304.3549}.

\bibitem[{\citenamefont{Fadin et~al.}(1975)\citenamefont{Fadin, Kuraev, and
  Lipatov}}]{Fadin:1975cb}
\bibinfo{author}{\bibfnamefont{V.~S.} \bibnamefont{Fadin}},
  \bibinfo{author}{\bibfnamefont{E.}~\bibnamefont{Kuraev}}, \bibnamefont{and}
  \bibinfo{author}{\bibfnamefont{L.}~\bibnamefont{Lipatov}},
  \bibinfo{journal}{Phys.Lett.} \textbf{\bibinfo{volume}{B60}},
  \bibinfo{pages}{50} (\bibinfo{year}{1975}).

\bibitem[{\citenamefont{Kuraev et~al.}(1977)\citenamefont{Kuraev, Lipatov, and
  Fadin}}]{Kuraev:1977fs}
\bibinfo{author}{\bibfnamefont{E.}~\bibnamefont{Kuraev}},
  \bibinfo{author}{\bibfnamefont{L.}~\bibnamefont{Lipatov}}, \bibnamefont{and}
  \bibinfo{author}{\bibfnamefont{V.~S.} \bibnamefont{Fadin}},
  \bibinfo{journal}{Sov.Phys.JETP} \textbf{\bibinfo{volume}{45}},
  \bibinfo{pages}{199} (\bibinfo{year}{1977}).

\bibitem[{\citenamefont{Balitsky and Lipatov}(1978)}]{Balitsky:1978ic}
\bibinfo{author}{\bibfnamefont{I.}~\bibnamefont{Balitsky}} \bibnamefont{and}
  \bibinfo{author}{\bibfnamefont{L.}~\bibnamefont{Lipatov}},
  \bibinfo{journal}{Sov.J.Nucl.Phys.} \textbf{\bibinfo{volume}{28}},
  \bibinfo{pages}{822} (\bibinfo{year}{1978}).

\bibitem[{\citenamefont{Kovchegov}(1999)}]{Kovchegov:1999yj}
\bibinfo{author}{\bibfnamefont{Y.~V.} \bibnamefont{Kovchegov}},
  \bibinfo{journal}{Phys.Rev.} \textbf{\bibinfo{volume}{D60}},
  \bibinfo{pages}{034008} (\bibinfo{year}{1999}), \eprint{hep-ph/9901281}.

\bibitem[{\citenamefont{Kovchegov}(2000)}]{Kovchegov:1999ua}
\bibinfo{author}{\bibfnamefont{Y.~V.} \bibnamefont{Kovchegov}},
  \bibinfo{journal}{Phys.Rev.} \textbf{\bibinfo{volume}{D61}},
  \bibinfo{pages}{074018} (\bibinfo{year}{2000}), \eprint{hep-ph/9905214}.

\bibitem[{\citenamefont{Balitsky}(1996)}]{Balitsky:1995ub}
\bibinfo{author}{\bibfnamefont{I.}~\bibnamefont{Balitsky}},
  \bibinfo{journal}{Nucl.Phys.} \textbf{\bibinfo{volume}{B463}},
  \bibinfo{pages}{99} (\bibinfo{year}{1996}), \eprint{hep-ph/9509348}.

\bibitem[{\citenamefont{Bahr et~al.}(2008)\citenamefont{Bahr, Gieseke, Gigg,
  Grellscheid, Hamilton et~al.}}]{Bahr:2008pv}
\bibinfo{author}{\bibfnamefont{M.}~\bibnamefont{Bahr}},
  \bibinfo{author}{\bibfnamefont{S.}~\bibnamefont{Gieseke}},
  \bibinfo{author}{\bibfnamefont{M.}~\bibnamefont{Gigg}},
  \bibinfo{author}{\bibfnamefont{D.}~\bibnamefont{Grellscheid}},
  \bibinfo{author}{\bibfnamefont{K.}~\bibnamefont{Hamilton}},
  \bibnamefont{et~al.}, \bibinfo{journal}{Eur.Phys.J.}
  \textbf{\bibinfo{volume}{C58}}, \bibinfo{pages}{639} (\bibinfo{year}{2008}),
  \eprint{0803.0883}.

\bibitem[{\citenamefont{Sjostrand et~al.}(2008)\citenamefont{Sjostrand, Mrenna,
  and Skands}}]{Sjostrand:2007gs}
\bibinfo{author}{\bibfnamefont{T.}~\bibnamefont{Sjostrand}},
  \bibinfo{author}{\bibfnamefont{S.}~\bibnamefont{Mrenna}}, \bibnamefont{and}
  \bibinfo{author}{\bibfnamefont{P.~Z.} \bibnamefont{Skands}},
  \bibinfo{journal}{Comput.Phys.Commun.} \textbf{\bibinfo{volume}{178}},
  \bibinfo{pages}{852} (\bibinfo{year}{2008}), \eprint{0710.3820}.

\bibitem[{\citenamefont{Collaboration}(2014)}]{CMS:2014oma}
\bibinfo{author}{\bibfnamefont{C.}~\bibnamefont{Collaboration}}
  (\bibinfo{collaboration}{CMS Collaboration}) (\bibinfo{year}{2014}),
  \bibinfo{note}{{CMS-PAS-FSQ-12-008}}.

\bibitem[{\citenamefont{Kwiecinski et~al.}(1996)\citenamefont{Kwiecinski,
  Martin, and Sutton}}]{Kwiecinski:1996td}
\bibinfo{author}{\bibfnamefont{J.}~\bibnamefont{Kwiecinski}},
  \bibinfo{author}{\bibfnamefont{A.~D.} \bibnamefont{Martin}},
  \bibnamefont{and} \bibinfo{author}{\bibfnamefont{P.}~\bibnamefont{Sutton}},
  \bibinfo{journal}{Z.Phys.} \textbf{\bibinfo{volume}{C71}},
  \bibinfo{pages}{585} (\bibinfo{year}{1996}), \eprint{hep-ph/9602320}.

\bibitem[{\citenamefont{Kwiecinski et~al.}(1997)\citenamefont{Kwiecinski,
  Martin, and Stasto}}]{Kwiecinski:1997ee}
\bibinfo{author}{\bibfnamefont{J.}~\bibnamefont{Kwiecinski}},
  \bibinfo{author}{\bibfnamefont{A.~D.} \bibnamefont{Martin}},
  \bibnamefont{and} \bibinfo{author}{\bibfnamefont{A.}~\bibnamefont{Stasto}},
  \bibinfo{journal}{Phys.Rev.} \textbf{\bibinfo{volume}{D56}},
  \bibinfo{pages}{3991} (\bibinfo{year}{1997}), \eprint{hep-ph/9703445}.

\bibitem[{\citenamefont{Sudakov}(1956)}]{Sudakov:1954sw}
\bibinfo{author}{\bibfnamefont{V.}~\bibnamefont{Sudakov}},
  \bibinfo{journal}{Sov.Phys.JETP} \textbf{\bibinfo{volume}{3}},
  \bibinfo{pages}{65} (\bibinfo{year}{1956}).

\bibitem[{\citenamefont{Catani et~al.}(1990{\natexlab{a}})\citenamefont{Catani,
  Fiorani, and Marchesini}}]{Catani:1989sg}
\bibinfo{author}{\bibfnamefont{S.}~\bibnamefont{Catani}},
  \bibinfo{author}{\bibfnamefont{F.}~\bibnamefont{Fiorani}}, \bibnamefont{and}
  \bibinfo{author}{\bibfnamefont{G.}~\bibnamefont{Marchesini}},
  \bibinfo{journal}{Nucl.Phys.} \textbf{\bibinfo{volume}{B336}},
  \bibinfo{pages}{18} (\bibinfo{year}{1990}{\natexlab{a}}).

\bibitem[{\citenamefont{Catani et~al.}(1990{\natexlab{b}})\citenamefont{Catani,
  Fiorani, and Marchesini}}]{Catani:1989yc}
\bibinfo{author}{\bibfnamefont{S.}~\bibnamefont{Catani}},
  \bibinfo{author}{\bibfnamefont{F.}~\bibnamefont{Fiorani}}, \bibnamefont{and}
  \bibinfo{author}{\bibfnamefont{G.}~\bibnamefont{Marchesini}},
  \bibinfo{journal}{Phys.Lett.} \textbf{\bibinfo{volume}{B234}},
  \bibinfo{pages}{339} (\bibinfo{year}{1990}{\natexlab{b}}).

\bibitem[{\citenamefont{Kimber et~al.}(2000)\citenamefont{Kimber, Martin, and
  Ryskin}}]{Kimber:1999xc}
\bibinfo{author}{\bibfnamefont{M.}~\bibnamefont{Kimber}},
  \bibinfo{author}{\bibfnamefont{A.~D.} \bibnamefont{Martin}},
  \bibnamefont{and} \bibinfo{author}{\bibfnamefont{M.}~\bibnamefont{Ryskin}},
  \bibinfo{journal}{Eur.Phys.J.} \textbf{\bibinfo{volume}{C12}},
  \bibinfo{pages}{655} (\bibinfo{year}{2000}), \eprint{hep-ph/9911379}.

\bibitem[{\citenamefont{Kutak et~al.}(2012)\citenamefont{Kutak, Golec-Biernat,
  Jadach, and Skrzypek}}]{Kutak:2011fu}
\bibinfo{author}{\bibfnamefont{K.}~\bibnamefont{Kutak}},
  \bibinfo{author}{\bibfnamefont{K.}~\bibnamefont{Golec-Biernat}},
  \bibinfo{author}{\bibfnamefont{S.}~\bibnamefont{Jadach}}, \bibnamefont{and}
  \bibinfo{author}{\bibfnamefont{M.}~\bibnamefont{Skrzypek}},
  \bibinfo{journal}{JHEP} \textbf{\bibinfo{volume}{1202}}, \bibinfo{pages}{117}
  (\bibinfo{year}{2012}), \eprint{1111.6928}.

\bibitem[{\citenamefont{Kutak}(2012)}]{Kutak:2012qk}
\bibinfo{author}{\bibfnamefont{K.}~\bibnamefont{Kutak}} (\bibinfo{year}{2012}),
  \eprint{1206.5757}.

\bibitem[{\citenamefont{Kimber et~al.}(2001)\citenamefont{Kimber, Martin, and
  Ryskin}}]{Kimber:2001sc}
\bibinfo{author}{\bibfnamefont{M.}~\bibnamefont{Kimber}},
  \bibinfo{author}{\bibfnamefont{A.~D.} \bibnamefont{Martin}},
  \bibnamefont{and} \bibinfo{author}{\bibfnamefont{M.}~\bibnamefont{Ryskin}},
  \bibinfo{journal}{Phys.Rev.} \textbf{\bibinfo{volume}{D63}},
  \bibinfo{pages}{114027} (\bibinfo{year}{2001}), \eprint{hep-ph/0101348}.

\bibitem[{\citenamefont{Mueller et~al.}(2013)\citenamefont{Mueller, Xiao, and
  Yuan}}]{Mueller:2012uf}
\bibinfo{author}{\bibfnamefont{A.}~\bibnamefont{Mueller}},
  \bibinfo{author}{\bibfnamefont{B.-W.} \bibnamefont{Xiao}}, \bibnamefont{and}
  \bibinfo{author}{\bibfnamefont{F.}~\bibnamefont{Yuan}},
  \bibinfo{journal}{Phys.Rev.Lett.} \textbf{\bibinfo{volume}{110}},
  \bibinfo{pages}{082301} (\bibinfo{year}{2013}), \eprint{1210.5792}.

\bibitem[{\citenamefont{Catani et~al.}(1991)\citenamefont{Catani, Ciafaloni,
  and Hautmann}}]{Catani:1990eg}
\bibinfo{author}{\bibfnamefont{S.}~\bibnamefont{Catani}},
  \bibinfo{author}{\bibfnamefont{M.}~\bibnamefont{Ciafaloni}},
  \bibnamefont{and} \bibinfo{author}{\bibfnamefont{F.}~\bibnamefont{Hautmann}},
  \bibinfo{journal}{Nucl.Phys.} \textbf{\bibinfo{volume}{B366}},
  \bibinfo{pages}{135} (\bibinfo{year}{1991}).

\bibitem[{\citenamefont{Catani and Hautmann}(1994)}]{Catani:1994sq}
\bibinfo{author}{\bibfnamefont{S.}~\bibnamefont{Catani}} \bibnamefont{and}
  \bibinfo{author}{\bibfnamefont{F.}~\bibnamefont{Hautmann}},
  \bibinfo{journal}{Nucl.Phys.} \textbf{\bibinfo{volume}{B427}},
  \bibinfo{pages}{475} (\bibinfo{year}{1994}), \eprint{hep-ph/9405388}.

\bibitem[{\citenamefont{Collins and Ellis}(1991)}]{Collins:1991ty}
\bibinfo{author}{\bibfnamefont{J.~C.} \bibnamefont{Collins}} \bibnamefont{and}
  \bibinfo{author}{\bibfnamefont{R.~K.} \bibnamefont{Ellis}},
  \bibinfo{journal}{Nucl. Phys.} \textbf{\bibinfo{volume}{B360}},
  \bibinfo{pages}{3} (\bibinfo{year}{1991}).

\bibitem[{\citenamefont{Lipatov}(1995)}]{Lipatov:1995pn}
\bibinfo{author}{\bibfnamefont{L.}~\bibnamefont{Lipatov}},
  \bibinfo{journal}{Nucl.Phys.} \textbf{\bibinfo{volume}{B452}},
  \bibinfo{pages}{369} (\bibinfo{year}{1995}), \eprint{hep-ph/9502308}.

\bibitem[{\citenamefont{Antonov et~al.}(2005)\citenamefont{Antonov, Lipatov,
  Kuraev, and Cherednikov}}]{Antonov:2004hh}
\bibinfo{author}{\bibfnamefont{E.}~\bibnamefont{Antonov}},
  \bibinfo{author}{\bibfnamefont{L.}~\bibnamefont{Lipatov}},
  \bibinfo{author}{\bibfnamefont{E.}~\bibnamefont{Kuraev}}, \bibnamefont{and}
  \bibinfo{author}{\bibfnamefont{I.}~\bibnamefont{Cherednikov}},
  \bibinfo{journal}{Nucl.Phys.} \textbf{\bibinfo{volume}{B721}},
  \bibinfo{pages}{111} (\bibinfo{year}{2005}), \eprint{hep-ph/0411185}.

\bibitem[{\citenamefont{van Hameren
  et~al.}(2013{\natexlab{a}})\citenamefont{van Hameren, Kotko, and
  Kutak}}]{vanHameren:2012if}
\bibinfo{author}{\bibfnamefont{A.}~\bibnamefont{van Hameren}},
  \bibinfo{author}{\bibfnamefont{P.}~\bibnamefont{Kotko}}, \bibnamefont{and}
  \bibinfo{author}{\bibfnamefont{K.}~\bibnamefont{Kutak}},
  \bibinfo{journal}{JHEP} \textbf{\bibinfo{volume}{1301}}, \bibinfo{pages}{078}
  (\bibinfo{year}{2013}{\natexlab{a}}), \eprint{1211.0961}.

\bibitem[{\citenamefont{van Hameren et~al.}(2012)\citenamefont{van Hameren,
  Kotko, and Kutak}}]{vanHameren:2012uj}
\bibinfo{author}{\bibfnamefont{A.}~\bibnamefont{van Hameren}},
  \bibinfo{author}{\bibfnamefont{P.}~\bibnamefont{Kotko}}, \bibnamefont{and}
  \bibinfo{author}{\bibfnamefont{K.}~\bibnamefont{Kutak}},
  \bibinfo{journal}{JHEP} \textbf{\bibinfo{volume}{1212}}, \bibinfo{pages}{029}
  (\bibinfo{year}{2012}), \eprint{1207.3332}.

\bibitem[{\citenamefont{van Hameren
  et~al.}(2013{\natexlab{b}})\citenamefont{van Hameren, Kutak, and
  Salwa}}]{vanHameren:2013csa}
\bibinfo{author}{\bibfnamefont{A.}~\bibnamefont{van Hameren}},
  \bibinfo{author}{\bibfnamefont{K.}~\bibnamefont{Kutak}}, \bibnamefont{and}
  \bibinfo{author}{\bibfnamefont{T.}~\bibnamefont{Salwa}}
  (\bibinfo{year}{2013}{\natexlab{b}}), \eprint{1308.2861}.

\bibitem[{\citenamefont{Kotko}(2014)}]{Kotko:2014aba}
\bibinfo{author}{\bibfnamefont{P.}~\bibnamefont{Kotko}} (\bibinfo{year}{2014}),
  \eprint{1403.4824}.

\bibitem[{\citenamefont{Dominguez et~al.}(2011)\citenamefont{Dominguez,
  Marquet, Xiao, and Yuan}}]{Dominguez:2011wm}
\bibinfo{author}{\bibfnamefont{F.}~\bibnamefont{Dominguez}},
  \bibinfo{author}{\bibfnamefont{C.}~\bibnamefont{Marquet}},
  \bibinfo{author}{\bibfnamefont{B.-W.} \bibnamefont{Xiao}}, \bibnamefont{and}
  \bibinfo{author}{\bibfnamefont{F.}~\bibnamefont{Yuan}},
  \bibinfo{journal}{Phys.Rev.} \textbf{\bibinfo{volume}{D83}},
  \bibinfo{pages}{105005} (\bibinfo{year}{2011}), \eprint{1101.0715}.

\bibitem[{\citenamefont{Xiao and Yuan}(2010)}]{Xiao:2010sp}
\bibinfo{author}{\bibfnamefont{B.-W.} \bibnamefont{Xiao}} \bibnamefont{and}
  \bibinfo{author}{\bibfnamefont{F.}~\bibnamefont{Yuan}},
  \bibinfo{journal}{Phys.Rev.Lett.} \textbf{\bibinfo{volume}{105}},
  \bibinfo{pages}{062001} (\bibinfo{year}{2010}), \eprint{1003.0482}.

\bibitem[{\citenamefont{Mulders and Rogers}(2011)}]{Mulders:2011zt}
\bibinfo{author}{\bibfnamefont{P.}~\bibnamefont{Mulders}} \bibnamefont{and}
  \bibinfo{author}{\bibfnamefont{T.}~\bibnamefont{Rogers}}
  (\bibinfo{year}{2011}), \eprint{1102.4569}.

\bibitem[{\citenamefont{Avsar}(2011)}]{Avsar:2011tz}
\bibinfo{author}{\bibfnamefont{E.}~\bibnamefont{Avsar}},
  \bibinfo{journal}{Int.J.Mod.Phys.Conf.Ser.} \textbf{\bibinfo{volume}{04}},
  \bibinfo{pages}{74} (\bibinfo{year}{2011}), \eprint{1108.1181}.

\bibitem[{\citenamefont{Avsar}(2012)}]{Avsar:2012hj}
\bibinfo{author}{\bibfnamefont{E.}~\bibnamefont{Avsar}} (\bibinfo{year}{2012}),
  \eprint{1203.1916}.

\bibitem[{\citenamefont{van Hameren
  et~al.}(2013{\natexlab{c}})\citenamefont{van Hameren, Kotko, and
  Kutak}}]{vanHameren:2013fla}
\bibinfo{author}{\bibfnamefont{A.}~\bibnamefont{van Hameren}},
  \bibinfo{author}{\bibfnamefont{P.}~\bibnamefont{Kotko}}, \bibnamefont{and}
  \bibinfo{author}{\bibfnamefont{K.}~\bibnamefont{Kutak}}
  (\bibinfo{year}{2013}{\natexlab{c}}), \eprint{1308.0452}.

\bibitem[{\citenamefont{Kutak and Kwiecinski}(2003)}]{Kutak:2003bd}
\bibinfo{author}{\bibfnamefont{K.}~\bibnamefont{Kutak}} \bibnamefont{and}
  \bibinfo{author}{\bibfnamefont{J.}~\bibnamefont{Kwiecinski}},
  \bibinfo{journal}{Eur.Phys.J.} \textbf{\bibinfo{volume}{C29}},
  \bibinfo{pages}{521} (\bibinfo{year}{2003}), \eprint{hep-ph/0303209}.

\bibitem[{\citenamefont{Martin et~al.}(2009)\citenamefont{Martin, Stirling,
  Thorne, and Watt}}]{Martin:2009iq}
\bibinfo{author}{\bibfnamefont{A.}~\bibnamefont{Martin}},
  \bibinfo{author}{\bibfnamefont{W.}~\bibnamefont{Stirling}},
  \bibinfo{author}{\bibfnamefont{R.}~\bibnamefont{Thorne}}, \bibnamefont{and}
  \bibinfo{author}{\bibfnamefont{G.}~\bibnamefont{Watt}},
  \bibinfo{journal}{Eur.Phys.J.} \textbf{\bibinfo{volume}{C63}},
  \bibinfo{pages}{189} (\bibinfo{year}{2009}), \eprint{0901.0002}.

\bibitem[{\citenamefont{Lai et~al.}(2010)\citenamefont{Lai, Guzzi, Huston, Li,
  Nadolsky et~al.}}]{Lai:2010vv}
\bibinfo{author}{\bibfnamefont{H.-L.} \bibnamefont{Lai}},
  \bibinfo{author}{\bibfnamefont{M.}~\bibnamefont{Guzzi}},
  \bibinfo{author}{\bibfnamefont{J.}~\bibnamefont{Huston}},
  \bibinfo{author}{\bibfnamefont{Z.}~\bibnamefont{Li}},
  \bibinfo{author}{\bibfnamefont{P.~M.} \bibnamefont{Nadolsky}},
  \bibnamefont{et~al.}, \bibinfo{journal}{Phys.Rev.}
  \textbf{\bibinfo{volume}{D82}}, \bibinfo{pages}{074024}
  (\bibinfo{year}{2010}), \eprint{1007.2241}.

\bibitem[{\citenamefont{Kotko}(2013)}]{Kotko_LxJet}
\bibinfo{author}{\bibfnamefont{P.}~\bibnamefont{Kotko}},
  \emph{\bibinfo{title}{{LxJet}}} (\bibinfo{year}{2013}), \bibinfo{note}{{C++}
  Monte Carlo program},
  \urlprefix\url{http://annapurna.ifj.edu.pl/~pkotko/LxJet.html}.

\bibitem[{\citenamefont{Sapeta}()}]{forward}
\bibinfo{author}{\bibfnamefont{S.}~\bibnamefont{Sapeta}},
  \emph{\bibinfo{title}{$\mathtt{forward}$}}, \bibinfo{note}{the code is
  available on request from the author}.

\bibitem[{\citenamefont{Jung et~al.}(2010)\citenamefont{Jung, Baranov, Deak,
  Grebenyuk, Hautmann et~al.}}]{Jung:2010si}
\bibinfo{author}{\bibfnamefont{H.}~\bibnamefont{Jung}},
  \bibinfo{author}{\bibfnamefont{S.}~\bibnamefont{Baranov}},
  \bibinfo{author}{\bibfnamefont{M.}~\bibnamefont{Deak}},
  \bibinfo{author}{\bibfnamefont{A.}~\bibnamefont{Grebenyuk}},
  \bibinfo{author}{\bibfnamefont{F.}~\bibnamefont{Hautmann}},
  \bibnamefont{et~al.}, \bibinfo{journal}{Eur.Phys.J.}
  \textbf{\bibinfo{volume}{C70}}, \bibinfo{pages}{1237} (\bibinfo{year}{2010}),
  \eprint{1008.0152}.

\end{thebibliography}

\end{document}